\documentclass[aps,prl,reprint,superscriptaddress,showpacs]{revtex4-1}
\usepackage[latin9]{inputenc}
\usepackage{color}
\usepackage{amsmath}
\usepackage{amssymb}
\usepackage{graphicx}

\makeatletter

\usepackage{eurosym}

\makeatother

\begin{document}
\title{Two stage Seebeck effect in charged colloidal suspensions}
\author{I.~Chikina}
\affiliation{LIONS, NIMBE, CEA, CNRS, Universit{\`e} Paris-Saclay, CEA Saclay,
91191 Gif-sur-Yvette, France}
\author{Sawako Nakamae}
\affiliation{Service de physique de l`etat condens{\`e}, SPEC, CEA, CNRS,
	Universit{\`e} Paris-Saclay, CEA Saclay, 91191 Gif-sur-Yvette, France}
\author{V.~Shikin}
\affiliation{ISSP, RAS, Chernogolovka, Moscow District, 142432 Russia}
\author{A.~Varlamov}
\affiliation{CNR-SPIN, c/o DICII-Universit{\`a} di Roma Tor Vergata, Via del Politecnico,
1, 00133 Roma, Italy}
\date{\today }

\begin{abstract}
	We discuss the peculiarities of the Seebeck effect in stabilized electrolytes containing the colloidal particles. Its unusual feature is the two stage character, with the linear increase of differential thermopower as the function of colloidal particles concentration $n_{\odot}$  during the first stage and  dramatic drop of it at small $n_{\odot}$ during the second one (steady state) \cite{Saco1}. We show that the properties of the initial state are governed by the  thermo-diffusion flows of the mobile ions of the stabilizing electrolyte medium itself and how the colloidal particles participate in formation of the electric field in the bulk of suspension. In its turn the specifics of the steady state in thermoelectric effect we attribute to considerable displacements of the massive colloidal particles in process of their slow thermal diffusion and break down of their electroneutrality in the vicinity of electrodes

\end{abstract}

\maketitle

\section{Introduction}

In recent years, liquid thermoelectric materials are emerging
as a cheaper alternative with semiconductor based solid counterpart
for low-grade waste heat recovery technologies. A breakthrough to
the enhancement of the thermo-electric efficiency of thermo-electrochemical cell has been achieved using ionic liquids \cite{dupont}. And more recently, the dispersion of charged colloidal particles (magnetic nanoparticle) was also found to increase the Seebeck coefficient of the host electrolyte. Incorporation
of the nano- and micro-meter sized colloidal particles can dramatically
change the transport properties of such systems. For example, in Ref.
\cite{Saco1} a novel use of charged colloidal solution was proposed
to improve the Seebeck coefficient of aqueous thermo-electrochemical
cell. The authors study transport properties of the charged colloidal
suspensions of iron oxide nanoparticles (maghemite) dispersed in aqueous
medium and report the values of the order of $1-1.5mV/K$ for the
Seebeck coefficient.
The inclusion of
Tetrabutyl ammonium as counterions, lead to an enhancement of the
fluids initial Seebeck coefficient by $15\%$ (at nanoparticle concentration
$1\%$). The authors of  Ref. \cite{AJ20} also indicate on high values of Seebeck coefficient ( $\approx 2 mV/K $) for many electrolyte-electrode combinations, what is much higher than existing prediction.

The charged colloidal suspensions of iron oxide nanoparticles (maghemite)
dispersed in aqueous medium was used by the authors of Ref. \cite{Saco1}.
Point in fact, that
when the colloidal particles are neutral, they can not exist stationary
in dilute solution, coagulating due to the van der Waals forces acting
between them. In order to prevent such coagulation processes, one
can immerse individual colloidal particles in the electrolyte specific
for each sort of them such that they acquire surface ions (e.g., hydroxyl groups, citrate, \textit{etc.}  \cite{Riedl,Bacri,Dubois}) resulting in accumulation by them	of very large structural charge $eZ$ ($|Z|\gg10$). Its sign can be both positive and negative, depending on the surface group type. Such procedure is called stabilization and
obtained suspension is considered stabilized.

\begin{figure}[!ht]
\includegraphics[width=1\linewidth]{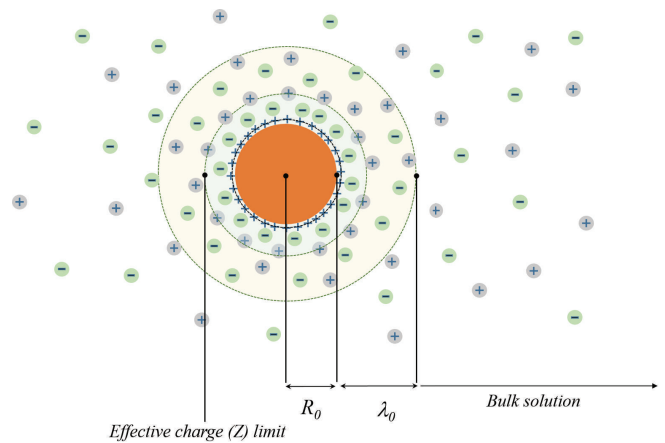} \caption{The schematic presentation of the multiply-charged colloidal particle
surrounded by the cloud of counter-ions.}
\label{Figure_1} 
\end{figure}

The large structural charge attracts counterions from the surrounding solvent
creating an electrostatic screening coat of the length $\lambda_{0}$
with an effective charge $-eZ$ (see Fig. \ref{Figure_1}). In these
conditions, nano-particles approaching within the distances
$r\le\lambda_{0}$ between them begin to repel each other preventing coagulation.
Corresponding theory of stabilized electrolyte was developed in Refs.
\cite{Derjaguin,LLSP,Verwey} and is often referred as the DLVO theory.

A clear manifestation of stabilization phenomenon occurs in such diluted
solutions especially clearly in the region of concentrations where
\begin{equation}
n_{\odot}\left(\lambda_{0}+R_{0}\right)^{3}\ll1,
\label{criterion}
\end{equation}
where $n_{\odot}$ is the density of colloidal particles and $R_{0}$
is the bare radius of the colloidal particle. Important, that in conditions
of Eq. (\ref{criterion}) validity the stabilized DLVO solution is
homogeneous.

In the case of thermocells considered here, with a temperature difference between two electrodes, the colloidal particles are dragged by the heat flow inducing a concentration gradient (Soret effect). The differences
in transport coefficients of various charged components of the stabilized
solution result in violation of the local electrical neutrality in
the latter when the temperature gradient is applied. Namely this circumstance
is the origin of the Seebeck effect occurring in the bulk of the stabilizing
electrolyte. It is why the thermo-diffusion displacements of the colloidal
particles can be observed only side by side with the Seebeck effect
in stabilized colloidal suspension. One can call this phenomenon as the electrostatic
potential difference induced by the Soret effect which is considered to be independent from the
redox reactions of the solutes in the electrolyte. At the same time, the concentration of the colloidal particles,
in accordance to the data of Ref. \cite{Saco1}, considerably effects
on the value of Seebeck coefficient.
Discussion of the possible reasons of this nontrivial effect is the
goal of presented work.

\section{The peculiarities of the Seebeck effect in colloidal solution}

When the temperature difference is imposed the two different kinds
of the diffusion flows occur in a thermocell. First, these are the thermo-diffusion
flows of the mobile ions of the stabilizing electrolyte, the second,
much slower, is that one of the colloidal particles. The diffusion
of positive and negative ions under the effect of temperature gradient
occurs with different rates what results in the appearance of the
charge separation and, as the consequence, formation of the internal
electric field, i.e. Seebeck effect.

\subsection{Initial state of Seebeck effect in colloidal solution}

In order to understand the effect of presence of the colloidal particles
presence in solution on the strength of the Seebeck effect at the initial state
let us assume that their concentration $n_{\odot}$ remains homogeneous.

The electric current in the conducting media in presence of electric
field and temperature gradient is described by the generic equation
\begin{equation}
{\bf j}=\sigma{\bf E}-\beta{\bf \nabla T}.\label{eq:current}
\end{equation}
Here $\beta=-S\sigma$ and $\sigma$ is electrical conductivity. For
the further convenience we expressed the former in terms of conductivity
and Seebeck coefficient (also called thermopower) $S$. Namely the
Seebeck coefficient $S$ determines the voltage appearing on the cell
with broken circuit related to the applied temperature difference
\begin{equation}
V=\int_{T_{1}}^{T_{2}}S(T)dT\label{eq:V}
\end{equation}
and it is this value which usually is measured in experiment (for
example in Ref. \cite{Saco1}).

Since the electrolyte consists of two oppositely charged subsystems
of positive and negative ions, its effective Seebeck coefficient is
determined by the sum of the coefficients $\beta_{\pm}$ of each ion
subsystem divided by the total conductivity ($\sigma_{+}+\sigma_{-}$)
of the solution: 
\begin{equation}
S_{\mathrm{tot}}=-\frac{\beta_{+}+\beta_{-}}{\sigma_{+}+\sigma_{-}}=\frac{S_{-}\sigma_{-}+S_{+}\sigma_{+}}{\sigma_{-}+\sigma_{+}}.\label{eq:sdiff}
\end{equation}

The conductivity of electrolyte containing some low enough concentration
(see condition (\ref{criterion})) of the colloidal particles was
recently studied in Ref. \cite{CSV20}, where the explicit expression
for it was obtained: 
\begin{eqnarray}
\sigma_{\mathrm{tot}}(n_{\odot}) & = & \sigma_{+}(n_{\odot})+\sigma_{-}(n_{\odot}),\nonumber \\
\sigma_{\pm}(n_{\odot})\! & = & \!\sigma_{\pm}^{(0)}\!\left[1\!+\!4\pi n_{\odot}\left(\frac{\gamma_{\pm}-1}{\gamma_{\pm}+2}\right)\left(R_{0}+\lambda_{0}\right)^{3}\right],\label{eq:sigmasimple}
\end{eqnarray}
where $\sigma_{\pm}^{(0)}$ are conductivities of the ion subsystems
in absence of colloidal particles, $\gamma_{\pm}=\sigma_{\odot}^{\pm}/\sigma_{\pm}^{(0)}$,
$\sigma_{\odot}^{\pm}$ is the effective conductivity of the screening
coat of the colloidal particle for the ions of corresponding sign,
while $R_{0}$ and $\lambda_{0}$ were introduced above. The effect
of screened colloidal particles on the conductivity realizing by the
positive ions, which can be drawn into the negatively charged screening
coat, is positive: $\gamma_{+}>1$. Vice versa, the conductivity carried
on by the negative ions is suppressed by presence of the colloidal
particles, since the screening coats of the latter repulse the former,
and corresponding $\gamma_{-}<1$. It turns out that the effect of
colloidal particles on the positive ions is dominant.

The conductivity growth as colloidal particles are introduced into
the solution one can be understood as the facilitation of charge transfer
in the media where some fraction of volume is occupied by these highly
conducting objects. In result, at the same intensity of electric field
$E$ current increases, i.e. conductivity growths. The situation is
different with the Seebeck coefficient $S$. It characterizes the
voltage response of the media on the applied temperature gradient
and there is no evident reasons to suppose the direct sensitivity
of $S_{\pm}$ to colloidal particles concentration.

In this assumption Eq. (\ref{eq:sdiff}) acquires form 
\begin{equation}
S_{\mathrm{tot}}(n_{\odot})=S_{\mathrm{tot}}^{(0)}+\Delta S(n_{\odot}),\label{eq:sdiff1}
\end{equation}
where 
\begin{equation}
S_{\mathrm{tot}}^{(0)}=\frac{S_{-}^{(0)}\sigma_{-}^{(0)}+S_{+}^{(0)}\sigma_{+}^{(0)}}{\sigma_{-}^{(0)}+\sigma_{+}^{(0)}},\label{eq:sdiff2}
\end{equation}
and 
\begin{eqnarray}
\Delta S_{\mathrm{tot}}(n_{\odot})= & \frac{12\pi n_{\odot}\left(\gamma_{+}-\gamma_{-}\right)(R_{0}\!+\!\lambda_{0})^{3}}{\left(\gamma_{+}+2\right)\left(\gamma_{-}+2\right)}\nonumber \\
 & \cdot\frac{\sigma_{-}^{(0)}\sigma_{+}^{(0)}\left(S_{+}^{(0)}-S_{-}^{(0)}\right)}{\left(\sigma_{-}^{(0)}+\sigma_{+}^{(0)}\right)^{2}}.\label{eq:sdiff3}
\end{eqnarray}
The measured in experiment Ref. \cite{Saco1} change of the Seebeck
coefficient as the function of colloidal particles concentration normalized
on its value in absence of the latter takes form

\begin{eqnarray}
\frac{\Delta S_{\mathrm{tot}}(n_{\odot})}{S_{\mathrm{tot}}^{(0)}}=12\pi n_{\odot}\frac{\left(\gamma_{+}-\gamma_{-}\right)(R_{0}\!+\!\lambda_{0})^{3}}{\left(\gamma_{+}+2\right)\left(\gamma_{-}+2\right)}\nonumber \\
\cdot\frac{\left(S_{+}^{(0)}-S_{-}^{(0)}\right)}{S_{-}^{(0)}\left(1+\sigma_{-}^{(0)}/\sigma_{+}^{(0)}\right)+S_{+}^{(0)}\left(1+\sigma_{+}^{(0)}/\sigma_{-}^{(0)}\right)}.\!\label{eq:sdiff4}
\end{eqnarray}

\begin{figure}[!ht]
\includegraphics[width=1\linewidth]{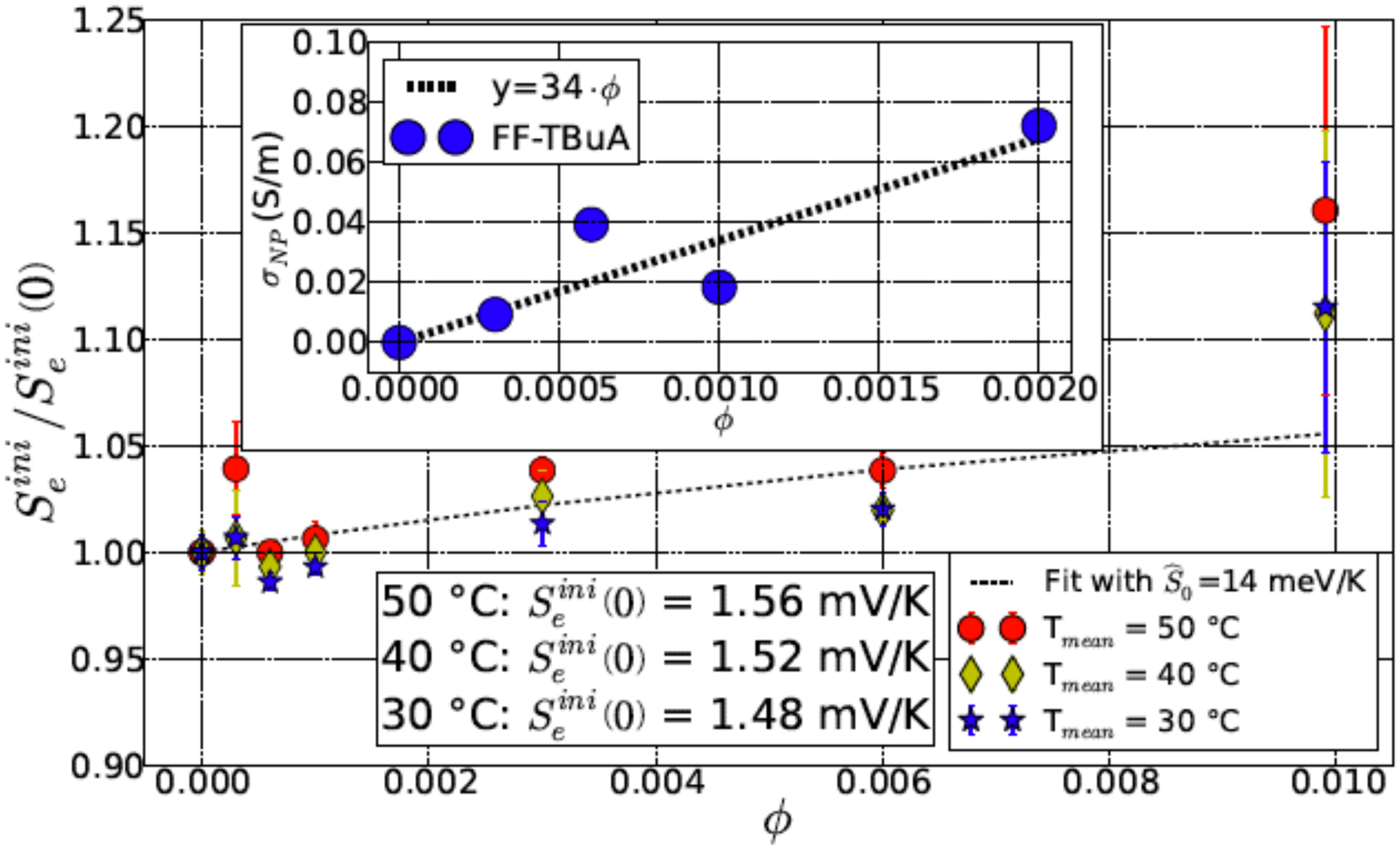} \caption{Normalized initial state Seebeck coefficient and electrical conductivity
(inset) as a function of nanoparticles concentration (taken from Ref.
\cite{Saco1}).}
\label{Seebeck_init} 
\end{figure}

It should be noted here that the thermogalvanic contribution to the overall temperature coefficient ($\Delta V/\Delta T$)  \cite{deBet} 
is not taken into consideration. This is justified because this term is additive to $ S_{\mathrm{tot}}^{(0)}$ and is independent of the nanoparticle concentration \cite{Saco1} and thus does not interfere with the $\Delta S_{\mathrm{tot}}^{(0)}$ in the Eq. (\ref{eq:sdiff4}).
The obtained result convincingly demonstrates, that the Seebeck coefficient
follows the linear growth of the colloidal particles concentration
in the considered range (\ref{criterion}) what corresponds to the
experimental findings (see Fig. \ref{Seebeck_init}). This linear
growth differs from that one of the colloidal solution conductivity
$\Delta\sigma_{\mathrm{tot}}(n_{\odot})$ (see Eq. (\ref{eq:sigmasimple}))
being proportional to the difference $S_{+}^{(0)}\!-\!S_{-}^{(0)}$
of the formal ion Seebeck coefficients in absence of colloidal particles.
Such result is very natural: Seebeck effect always (in metals, in
semiconductors) is related to the asymmetry of the charge carriers.

Hence, the direct measurements of $\sigma_{\mathrm{tot}}(0),$ $S_{\mathrm{tot}}^{(0)}$
and the values of the slopes in the dependencies of conductivity and
Seebeck effect as the function of colloidal particles concentration,
side by side with the independent knowledge of $\gamma_{\pm}$ and
$R_{0}+\lambda_{0}$ (according to Refs. \cite{CSV20,Saco1} $\lambda_{0} \approx 60 \AA, R_{0} \approx 70 \AA$)  
allow to determine the values $\sigma_{\pm}^{(0)}$.

\subsection{Steady state of Seebeck effect in colloidal solution}

Application of the temperature difference across the thermocell
results in charge separation among electrolyte ions. This happens
first due to the difference of their coefficients $\beta_{\pm}$,
second due to the difference of their diffusion coefficients \cite{CSV12}.
The latter contribution to the Seebeck coefficient is specific for
semiconductors and electrolytes and accounts for thermodiffusion of
charged particles, which occurs in presence of temperature gradient.
In result, at the initial state of the Seebeck effect, the accumulative
layers of ions of opposite charges are formed in the vicinity of electrodes.

\begin{figure}[!ht]
\includegraphics[width=1.1\linewidth]{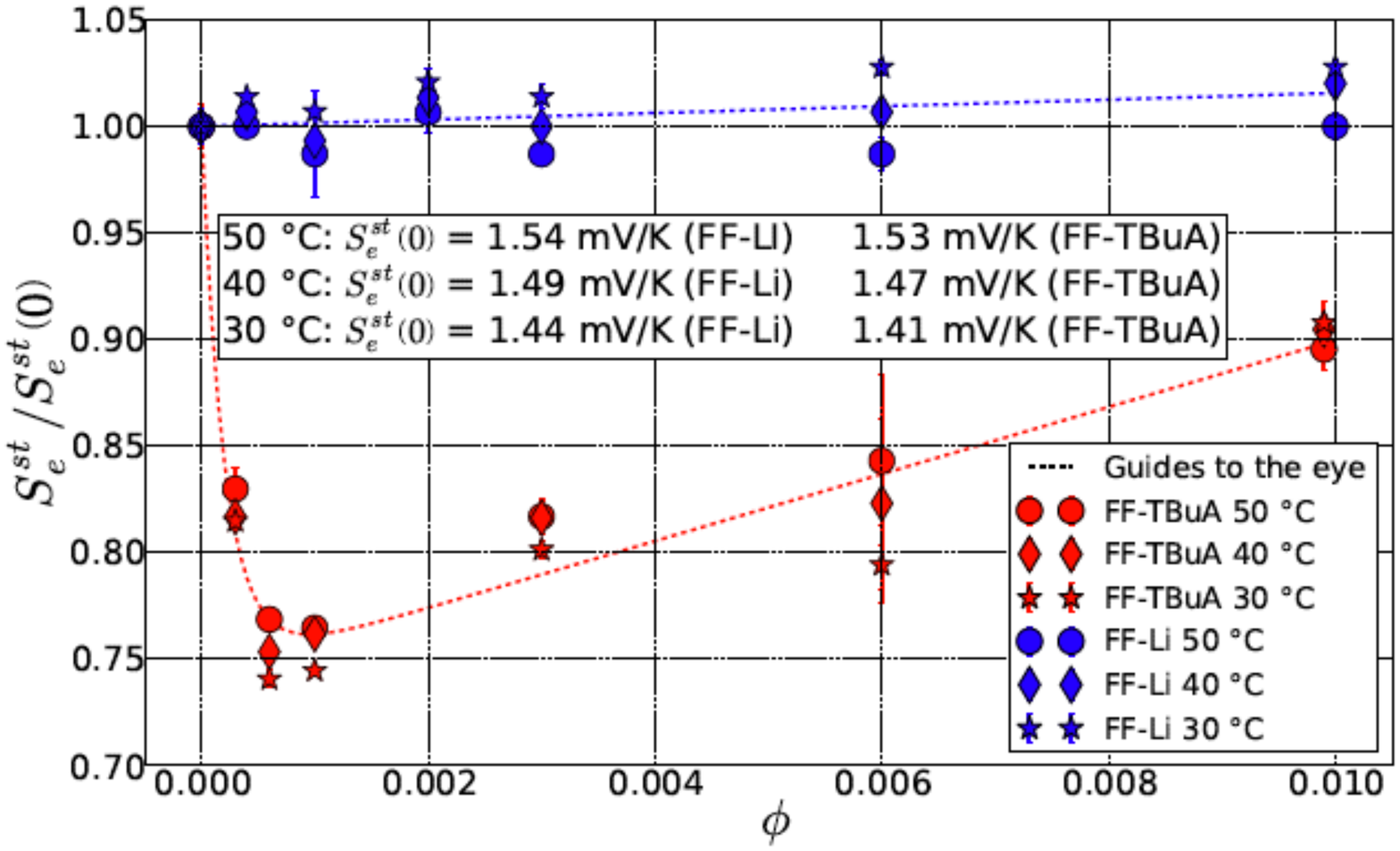} \caption{Normalized steady state Seebeck coefficient as a function of nanoparticles
concentration (taken from Ref. \cite{Saco1}).}
\label{seebeck_min} 
\end{figure}

When the electrolyte contains some relatively small concentration
of colloidal particles in process of time the measured value of Seebeck
coefficient decreases. This decrease as the function of nano-particles
concentration is drastic at the beginning, then Seebeck coefficient
reaches the minimum, and finally, in accordance to Eq. (\ref{eq:sdiff4}),
it linearly grows (see Fig. \ref{seebeck_min}) \cite{Saco1}. 

One of the reasons of its occurrence can be the complex structure
of the colloidal particles surrounded by their screening coatings. The
process of thermodiffusion results in their slow drift whose direction depends sensitively on the ionic environment surrounding the colloidal particles.
Let us recall, that in Ref. \cite{CSV20}
the colloidal particle screening was considered in the spherically
symmetrical situation by means of solution of the Poisson equation
with the zero boundary conditions at infinity. Close to the electrode
the Seebeck electric field in electrolyte is formed mainly due to
the redistribution of ions and becomes non-homogeneous \cite{CSV15}.
As the consequence, the colloidal complexes, acquiring induced dipole
moment, get pulled into the domain of higher fields.

In the vicinity of the thermocell extremities the difference in electrostatic
attraction of the charged counterions of the coat and the nucleos
of the colloidal particle to the metallic electrode enters in play.
 \textcolor{green}Indeed, such attraction forces turn out to be very
different for weakly charged counterions of the coating layer and strongly
charged ($Q=Ze$) core of the colloid. In result, the clusters lose
their electro-neutrality and start to compensate the charge of ionic
accumulative layer.

Let us evaluate the value of nanoparticles concentration corresponding
to the minimum of Seebeck coefficient at Fig. \ref{seebeck_min}.
For this purpose to do this we recall some facts of electrostatics. The
problem of a point charge interaction with the conducting plane, separating
two semi-spaces with different dielectric constants $\epsilon_{1}$
and $\epsilon_{2}$, can be reduced to that one of the charge interaction
with the corresponding electrostatic image charge behind the plane (see Ref.
\cite{LLESS}): 
\begin{equation}
F_{\epsilon}(z)=\frac{Q^{2}(\epsilon_{1}-\epsilon_{2})}{4\epsilon_{1}(\epsilon_{1}+\epsilon_{2})z^{2}}.\label{force}
\end{equation}
Here the charge is supposed to be placed at the distance $z$ from
the plane in the semi-space with dielectric constant $\epsilon_{1}$.

When the semi-space is filled by electrolyte the electrostatic image
force (\ref{force}) is screened at the distances of the order of
Debye length from the plane (Refs. \cite{Wagner,Onsager}): 
\begin{equation}
F_{\mathrm{WO}}(z)=F_{\epsilon}(z)\exp{\left(-\frac{2z}{\lambda_{0}}\right)}.\label{forceWO}
\end{equation}
with corresponding potential 
\[
U_{\mathrm{WO}}(z)\!=\!-\!\int_{z}^{\infty}\!\!F_{\mathrm{WO}}(x)dx\!=\!\frac{Q^{2}(\epsilon_{1}\!-\!\epsilon_{2})}{2\lambda_{0}\epsilon_{1}(\epsilon_{1}\!+\!\epsilon_{2})}\Gamma\left(\!-1,\frac{2z}{\lambda_{0}}\right),
\]
whereas $\Gamma\left(s,x\right)$ is the upper incomplete gamma function.
In other words, a charged particle located in the electrolyte at distances
exceeding the Debye length $\lambda_{0}$ from the electrode interacts
exponentially weakly with it. In the case under consideration we assume
the dielectric constant of the metallic electrode $\epsilon_{2}\rightarrow\infty$,
while $\epsilon_{1}=\epsilon_{aq}$. The corresponding electrostatic
energy for the the nucleos of the colloidal particle is 
\begin{equation}
U_{\mathrm{WO}}(z)=-\frac{Z^{2}e^{2}}{2\lambda_{0}\epsilon_{aq}}\Gamma\left(-1,\frac{2z}{\lambda_{0}}\right).\label{eq:WOenergy}
\end{equation}
\begin{figure}[!ht]
\includegraphics[width=1\linewidth]{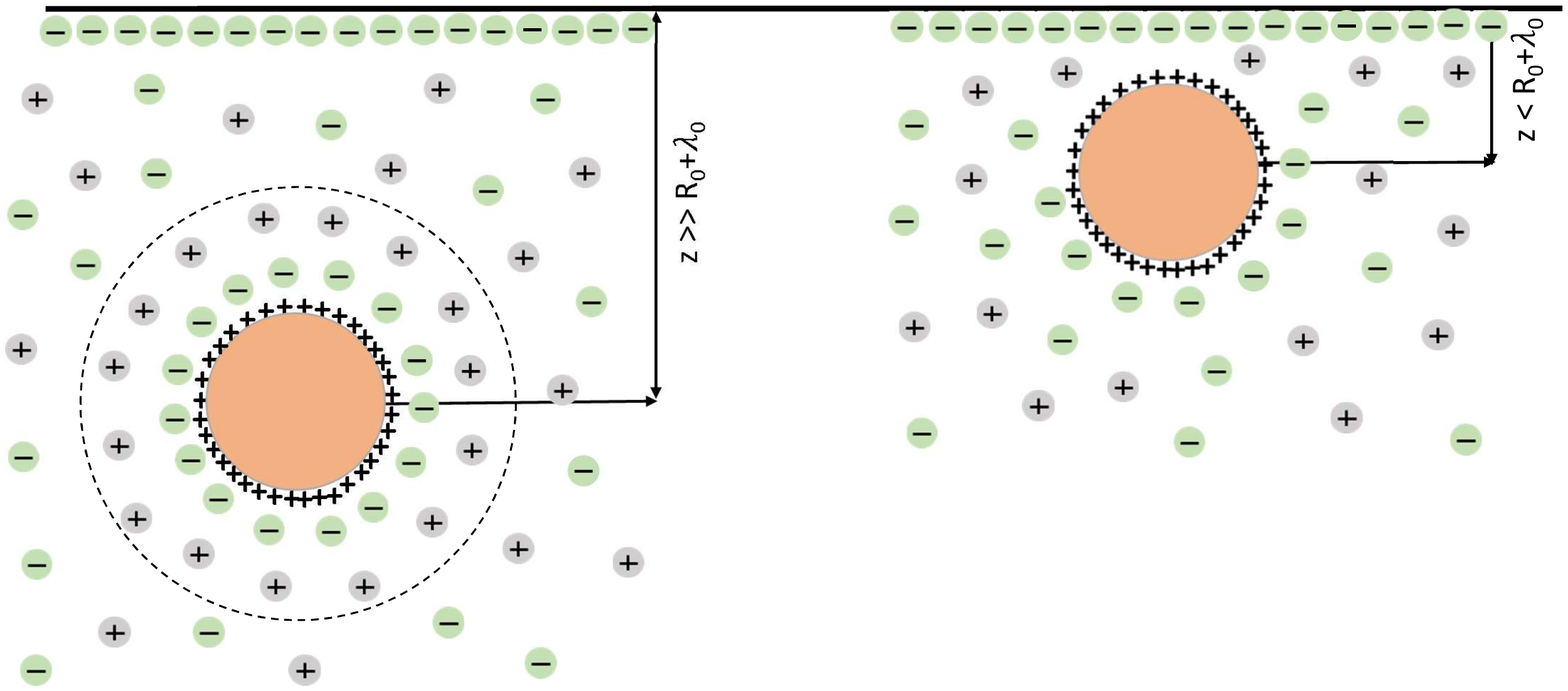} \caption{Until the colloidal particle size exceeds the Debye length it keeps
its integrity (a). When the colloidal particle approaches the charged
electrode at the distances less than its size the cluster loses its
electro-neutrality (b).}
\label{figure_4} 
\end{figure}

The effective radius of the colloidal particle is the sum of the radius
of charged nucleous and the thickness of the screening coat: $R_{0}+\lambda_{0}$.
Hence until its size exceeds the Debye length the colloidal particle
keeps its integrity (see Figure \ref{figure_4}a) and the potential
energy of its nucleous interaction with the electrode is determined
by Eq. (\ref{eq:WOenergy}). When the colloidal particle approaches
the charged electrode at the distances less than its size it
loses its electro-neutrality (see Figure \ref{figure_4}b) and its
consideration in the framework of Poisson equation with infinite boundary
conditions \cite{CSV20} is no longer applicable. The detailed study
of electrostatic interaction between a planar surface and a charged
sphere immersed in the electrolyte media was performed in Ref. \cite{ohshima}.
The author found corresponding energy $U_{\mathrm{HO}}$ of such interaction in the form
\begin{equation}
U_{\mathrm{HO}}(z)=U_{0}exp\left(-\frac{z-R_{0}}{\lambda_{0}}\right).\label{eq:HOenr}
\end{equation}

Applying this consideration to the case of the colloidal
particle core and matching Eqs. (\ref{eq:WOenergy})-(\ref{eq:HOenr})
at the distance $z=R_{0}+\lambda_{0}$ one can find the constant $U_{0}$:
\[
U_{0}\sim-\frac{Z^{2}e^{2}}{\lambda_{0}\epsilon_{aq}}\Gamma\left(-1,2+\frac{2R_{0}}{\lambda_{0}}\right).
\]

Now one can estimate the maximal colloidal particles surface concentration
$N_{\odot}^{max}$ which can be localized in the vicinity of the electrode
by means of the discussed above mechanism. Their attraction to the
electrode, charged due to the presence of excess ions with the surface
concentration $N_{-}$, continue until the latter will not be compensated
by the positive charges of the colloidal nuclei: 
\begin{equation}
N_{\odot}^{max}=N_{-}/Z.\label{Ncrit}
\end{equation}
The value $N_{-}$ can be found knowing the values of Seebeck coefficient
at the end of Initial state $S_{\mathrm{tot}}$, and the temperature
gradient. Indeed, considering the thermocell with electrolyte as the
parallel plate capacitor one can write: 
\[
E=\frac{4\pi eN_{-}}{\epsilon_{aq}},
\]
where from 
\begin{equation}
N_{-}=\frac{\epsilon_{aq}}{4\pi e}S_{\mathrm{tot}}\left(\frac{\varDelta T}{\varDelta L}\right).\label{eq:Ecr}
\end{equation}

What concerns the value of the surface concentration of the colloidal particles $N_{\odot}$ it can found by comparison
between the homogeneous distribution of the colloidal particles and that one in the presence of electrostatic potential of the electrode, determined
by Eqs.(\ref{eq:WOenergy})-(\ref{eq:HOenr}): 
\begin{widetext}
\begin{equation}
N_{\odot}\!=\!n_{\odot}\left\{ \int_{R_{0}}^{R_{0}\!+\!\lambda_{0}}\left(exp\left[\frac{Z^{2}e^{2}\Gamma\left(\!-1,2\!+\!\frac{2R_{0}}{\lambda_{0}}\right)}{\lambda_{0}k_{B}T\epsilon_{aq}}exp\left(\!-\!\frac{z-R_{0}}{\lambda_{0}}\right)\right]\!-1\!\right)dz\!+\!\int_{R_{0}\!+\!\lambda_{0}}^{\infty}\left(exp\left[\frac{Z^{2}e^{2}}{2\lambda_{0}k_{B}T\epsilon_{aq}}\Gamma\left(\!-\!1,\frac{2z}{\lambda_{0}}\right)\right]\!-\!1\right)dz\right\} .\label{eq:20}
\end{equation}
\end{widetext}
This equation relates the surface concentration of the colloidal particles $N_{\odot}$ with their volume concentration $n_{\odot}$.

Using the asymptotic expressions for incomplete Gamma-function 
\[
\Gamma\left(-1,x\right)=\left\{ \begin{array}{cc}
e^{-x}/x^{2}, & x\gg1\\
1/x & x\ll1
\end{array}\right.
\]
and making sure that $Z^{2}e^{2}\ll\lambda_{0}k_{B}T\epsilon_{aq}$
( the characteristic values of the parameters $\text{Z}\approx300,\epsilon_{aq}\approx80$,
$\lambda_{0} \approx 60 \AA, R_{0} \approx 70 \AA$) one
can find analyzing the date of Ref. \cite{Saco1} with the help of
Ref. \cite{CSV20}). Hence, the exponents in Eq. (\ref{eq:20}) can
be expanded, what results in 
\begin{equation}
N_{\odot}\sim n_{\odot}\frac{Z^{2}e^{2}}{k_{B}T\epsilon_{aq}}\frac{e^{-2\left(1+\frac{R_{0}}{\lambda_{0}}\right)}}{\left(1+\frac{R_{0}}{\lambda_{0}}\right)^{2}}.\label{eq:ncr}
\end{equation}

In order to estimate the values let us express Eq. (\ref{eq:ncr})
in terms of the Rydberg unit of energy $\mathrm{Ry}=e^{2}/a_{B}\textrm{=13.6 eV}$
($a_{B}=0.53\textrm{\AA}$ is the Bohr radius): 
\[
N_{\odot}\approx\left(n_{\odot}\lambda_{0}\right)\frac{Z^{2}\mathrm{Ry}}{2k_{B}T\epsilon_{aq}}\left(\frac{a_{B}}{\lambda_{0}}\right)\frac{e^{-2\left(1+\frac{R_{0}}{\lambda_{0}}\right)}}{\left(1+\frac{R_{0}}{\lambda_{0}}\right)^{2}}\sim10^{-2}\left(n_{\odot}\lambda_{0}\right).
\]
Returning to Eq. (\ref{Ncrit}) and substituting in it Eq. (\ref{eq:ncr})
one finds 
\begin{eqnarray}
n_{\odot}^{max}=\frac{\epsilon_{aq}^{2}S_{\mathrm{tot}}}{2\pi eZ^{3}a_{B}}\left(\frac{\varDelta T}{\varDelta L}\right)\left(\frac{k_{B}T}{\mathrm{Ry}}\right)\nonumber \\
\cdot e^{2\left(1+\frac{R_{0}}{\lambda_{0}}\right)}\left(1+\frac{R_{0}}{\lambda_{0}}\right)^{2}\sim10^{15}cm^{-3}.\label{eq:nfin}
\end{eqnarray}

In the dimensionless units of Ref. \cite{Saco1} the maximal concentration
of colloidal particles determined by Eq. (\ref{eq:nfin}) corresponds
to $\phi^{min}\le0.001$ (see Fig. \ref{seebeck_min}). Since we know
that in these units $\phi=0.006\rightarrow n_{\odot}\simeq5.45\cdot10^{15}cm^{-3}$,
it is easy to recalculate that the estimation (\ref{eq:nfin}) $n_{\odot}^{max}\sim10^{15}cm^{-3}\rightarrow\phi^{min}\sim0.01$,
which surprisingly well coincides to the experimental findings of Ref.
\cite{Saco1} (see Fig. \ref{seebeck_min}) considering the imperfect nature of metallic electrodes used in the \textit{real} thermocells.

\section{Discussion}

In this work we have studied the nontrivial role of colloidal particles
in formation of the Seebeck field in charged colloidal suspension. The reasons
for the two-stage character of the Seebeck effect observed in stabilized
colloidal electrolytes are discussed. It is shown that the ``Initial
state'' is related to the phenomenon of thermal diffusion of the
ions of the stabilizing electrolyte itself. The ensuing ``Steady
state'' occurs when the thermodiffusion displacement of the colloidal
particles becomes essential.

We demonstrate that, surprisingly, the neutral colloids affect on
the Seebeck coefficient already in the ``initial state''. This happens
due to their influence on the polyelectrolyte conductivity. As it
was shown in Ref. \cite{CSV20} the presence in the bulk of stabilizing
electrolyte of rarefied gas of colloids having a relatively large
conductivity of the screening coats increases its effective conductivity.
Accounting for this fact appears to explain the linear growth of the Seebeck coefficient as a function of the colloidal particles concentration observed in the experiment.

The observed sharp drop of the Seebeck coefficient when the small
concentration of the colloidal particles is added to stabilizing electrolyte
\cite{Saco1} is noteworthy. We propose the explanation of this feature
basing on the specific behavior of colloidal particle in the vicinity
of electrode. Approaching the latter colloid loses its neutrality,
discharging the accumulative layer of ions formed during the Initial
state. The decrease of the accumulative layer charge results in the
drop of Seebeck signal. Our qualitative estimations give surprisingly
good correspondence to experimental findings.

Finally, one can shed light on the discrepancy between the steady
state establishment time lapse in the experiment ($\tau_{exp}\sim8$ hours)
and the theoretical estimations counterpart by the authors of Ref.
\cite{Saco1}. The latter, $\tau_{theor}=(\Delta L)^{2}/\mathcal{D}_{\odot}\sim100$
hours is based on the entire length scale of the thermocell. In the model developed here, the formation of mirror charges occurs in the close vicinity of the electrode/electrolyte interfaces. Furthermore,
we want
to attract attention to the fact that, due to the inhomogeneity of
the charge density distribution along the thermocell length, the electric field
 also becomes non-homogeneous \cite{CSV12}, what results in polarization
of the colloidal particles and acceleration of their motion with respect
to a simple diffusion.

\section*{Acknowledgments}

This work is supported by European Union's Horizon 2020 research and
innovation program under the grant agreement n 731976 (MAGENTA).

\end{document}